\documentclass[aps,prd]{revtex4}

\usepackage{amsmath}
\usepackage{amssymb}
\usepackage{bm}
\usepackage{graphicx}
\usepackage{multirow}

\usepackage{textcomp}

\allowdisplaybreaks[1]

\begin{document}

\title{Understanding small neutrino mass and its implication}
\author{Hsiang-nan Li}
\affiliation{Institute of Physics, Academia Sinica,
Taipei, Taiwan 115, Republic of China}

\date{\today}

\begin{abstract}
We have derived previously the relations between the neutrino masses and mixing angles 
in a dispersive analysis on the mixing of neutral leptonic states. The only involved 
assumption is that the electroweak symmetry of the Standard Model (SM) is restored at 
a high energy scale in some new physics scenario, which diminishes the box diagrams 
responsible for the mixing. Here we include corrections to the analysis up to three 
loops, arising from exchanges of additional neutral and charged scalars in the electroweak 
symmetric phase. The solution to the dispersion relation for the $\mu^-e^+$-$\mu^+e^-$ 
mixing generates a typical neutrino mass $m_\nu\sim O(1)$ eV in the SM unambiguously. The 
solution also favors the normal ordering of the neutrino masses over the inverted one, and 
links the large electroweak symmetry restoration, i.e., new physics scale to the small 
neutrino mass. 
\end{abstract}

\maketitle

\section{INTRODUCTION}


It has been a long pursuit in particle physics to understand the flavor structure of the 
Standard Model (SM). We proposed recently \cite{Li:2023ncg,Li:2023dqi,Li:2023yay} that 
analyticity imposes additional dynamical constraints on the SM parameters beyond the Lagrangian 
level, which can account for the mass hierarchy and the distinct mixing patterns between quarks 
and leptons. Take the mixing of the neutral leptonic states 
$\mu^- e^+$ and $\mu^+ e^-$ as an example, where the muon is treated as being heavy with the 
invariant mass squared $s$, and the electron is regarded as being massless. The mixing amplitude 
governed by the leading-order (LO) box diagrams must respect a dispersion relation formulated in 
the $s$ complex plane. It was argued that the mixing phenomenon will disappear, if the 
electroweak symmetry of the SM is restored at a high energy scale $\Lambda$ 
\cite{Chien:2018ohd,Huang:2020iya}. The composite Higgs model described in \cite{Kaplan:1983fs} 
provides a suitable platform for the argumentation; the electroweak group in their model is 
broken at a scale much lower than the condensate scale, implying the existence of a symmetric 
phase which we refer to. The disappearance of the mixing at $s>\Lambda^2$ was taken as the LO 
input to the dispersion relation, and the corresponding solution at low $s<\Lambda^2$, i.e., 
in the symmetry broken phase, was found to effectively bind the neutrino masses and the 
Pontecorvo–Maki–Nakagawa–Sakata (PMNS) matrix elements involved in the box diagrams.

Several important observations were extracted from the dispersive constraints on the 
neutrino masses and the PMNS matrix elements \cite{Li:2023ncg}. The neutrino mass ordering, 
whose various scenarios have not been discriminated experimentally, remains unsettled in 
neutrino physics \cite{PDG}. It was noticed that the neutrino masses in the normal ordering 
(NO), instead of in the inverted ordering (IO), match the observed PMNS matrix elements. The 
neutrino mixing angles larger than the quark ones are attributed to the inequality of the 
mass ratios, $m_2^2/m_3^2\gg m_s^2/m_b^2$, where $m_2$ ($m_3$) is the second-generation 
(third-generation) neutrino mass and $m_s$ ($m_b$) is the $s$ ($b$) quark mass. The muon in 
the box diagrams can be replaced by another heavy lepton $\tau$, so the $\tau^-e^+$-$\tau^+e^-$ 
mixing involves the same intermediate neutrino states, and obeys the same constraints. 
The $\mu$-$\tau$ reflection symmetry \cite{Harrison:2002et} was thus realized easily; the 
combined dispersive analyses of the $\mu^-e^+$-$\mu^+e^-$ and $\tau^-e^+$-$\tau^+e^-$ mixings 
demands unequivocally the mixing angle $\theta_{23}\approx 45^\circ$ in agreement with its 
measured value around the maximal mixing. We emphasize that the above observations were obtained 
without resorting to specific new ingredients beyond the SM (BSM), contrary to conventional 
endeavors on this topic (for recent publications, see \cite{Alvarado:2020lcz,Xu:2023kfi,
Patel:2023qtw,Bora:2023teg,Thapa:2023fxu,Chung:2023rie,Supanyo:2023jkh,Lampe:2024gqo,
Shaikh:2024ufv}). The only required assumption is the restoration of the electroweak symmetry 
at a high energy scale. The successful explanation on part of the SM flavor structure
then also sheds light on potential new physics models.

Precisely speaking, our previous study explored the connection between the mixing angles and 
the ratios of the neutrino masses, instead of the absolute values of the neutrino masses 
\cite{Bilenky:2002aw}. Therefore, one of the key issues in neutrino physics, i.e., the 
smallness of the neutrino masses, has not been addressed. As a follow-up, we include 
loop corrections to the LO formalism, which arise from exchanges of additional
neutral and charged scalars in the electroweak symmetric phase.  The solution to 
the dispersion relation for the $\mu^-e^+$-$\mu^+e^-$ mixing with the high-energy inputs
up to three loops generates a tiny neutrino mass scale in the SM via the equation
\begin{eqnarray}
m_\nu^2\sqrt{\ln\frac{m_W^2}{m_\nu^2}}\approx \frac{\sqrt{3}}{128\pi^2}\frac{m_\mu^2 m_e^2}{v^2},
\label{mn3}
\end{eqnarray}
with the $W$ boson mass $m_W$, the muon (electron) mass $m_\mu$ ($m_e$) and the vacuum 
expectation value (VEV) $v$ of the Higgs field. Compared to the well-known seesaw mechanism 
\cite{see}, the VEV $v$ plays the role of a heavy Majorana neutrino mass in suppressing 
neutrino masses relative to charged lepton ones. The coefficient on the right-hand side,
resulting from the measure of a three-loop integral, provides additional suppression. 
Equation~(\ref{mn3}) leads to $m_\nu\sim O(1)$ eV unambiguously, which, as an 
order-of-magnitude estimate, is compatible with the upper bound on the neutrino mass 
$m_\nu<0.9$ eV at 90\% CL measured by the KATRIN Collaboration via the the endpoint spectrum 
of the tritium $\beta$-decay \cite{Nature}. 

We also find from the solution that the tiny neutrino mass hints a large electroweak 
symmetry restoration scale $\Lambda$ through the approximate formula
\begin{eqnarray}
\ln\frac{\Lambda}{m_W}\sim O(1)\sqrt{\ln\frac{m_W^2}{m_\nu^2}},\label{mn32}
\end{eqnarray}
where $O(1)$ denotes a coefficient of order of unity; the small neutrino mass is indeed 
linked to a new physics scale \cite{Mohapatra:2006gs}. As mentioned before, the neutrino 
masses in the NO satisfy better the LO dispersive constraints from the mixing of neutral 
leptonic states than in the IO \cite{Li:2023ncg}. We examine whether this 
postulation sustains its validity by deriving the constraint under the loop corrections,
\begin{eqnarray}
U^*_{\mu 1} U_{e1}(m_3-m_1)+U^*_{\mu 2} U_{e2}(m_3-m_2)=O\left(\frac{m_\nu^2}{v}\right)
\approx 0,\label{mn33}
\end{eqnarray}
with the PMNS matrix elements $U_{\mu i}$ and $U_{ei}$ and the first-generation neutrino mass
$m_1$. It will be demonstrated that Eq.~(\ref{mn33}) confirms the preference on the fitted 
parameters associated with the NO. The discrimination is mainly due to the different fitted $CP$ 
phases between the two possible mass orderings.

The rest of the paper is organized as follows. We recapture our framework \cite{Li:2023ncg}
in Sec.~II, and present the real part of the mixing amplitude in the symmetric phase and the 
imaginary part in the broken phase. The two pieces are substituted into the
dispersion relation respected by the mixing amplitude in Sec.~III to construct the
solution. Equations~(\ref{mn3})-(\ref{mn33}) are established, and their physical implications
are elaborated. Section~IV contains the conclusion and outlook. The real
part of the mixing amplitude is evaluated up to three loops in the Appendix.

\section{FORMALISM}

Consider the mixing of the neutral leptonic states $\mathcal{L}^-_L\ell^+_L$ and 
$\mathcal{L}^+_L\ell^-_L$, where $\mathcal{L}$ ($\ell$) stands for a massive (light) charged 
lepton and $L$ labels the left-handedness. The right-handedness will be labeled by $R$ below. 
Before the electroweak symmetry breaking, all particles are massless, and leptons are in their 
flavor eigenstates. The mixing occurs through exchanges of charged or neutral scalars among 
leptons, whose strengths are characterize by the Yukawa couplings. The Yukawa matrix elements 
are not all independent, so the discussion of the mixing in the symmetric phase will be more 
tedious, if it is based on the Yukawa matrices. A more transparent picture is attained by 
implementing the fermion field transformations adopted in the symmetry broken phase. The 
standard unitary transformation of fermion fields from the flavor eigenstates with the 
superscript $f$ to the mass eigenstates reads
\begin{eqnarray}
\nu_L^{(f)}= U_\nu\nu_L,\;\;\;\;\nu_R^{(f)}= V_\nu\nu_R,\;\;\;\;
\ell_L^{(f)}= U_\ell \ell_L,\;\;\;\;\ell_R^{(f)}= V_\ell \ell_R,
\end{eqnarray}
which diagonalizes the Yukawa matrices with the superscript $d$
\begin{eqnarray}
Y_\nu^{(d)}=U_\nu^\dagger Y_\nu V_\nu,\;\;\;\;Y_\ell^{(d)}=U_\ell^\dagger Y_\ell V_\ell,
\end{eqnarray}
and defines the PMNS matrix
\begin{eqnarray}
U= U_\ell^\dagger  U_\nu.\label{pm}
\end{eqnarray}

The relevant Lagrangian in the symmetric phase is then given by  
\begin{eqnarray}
Y_\nu^{(d)}\bar\nu_L(-\bar\phi^0)\nu_R
+Y_\ell^{(d)}\bar\ell_L\phi^0\ell_R
+Y_\nu^{(d)}\bar\ell^{\prime}_L\phi^-\nu_R
+Y_\ell^{(d)}\bar\nu^{\prime}_L\phi^+\ell_R+h.c.,\label{la}
\end{eqnarray}
after the fermion field transformation,
where the real component of $\phi^0$ in the first two terms corresponds to the SM Higgs 
boson. The imaginary component of $\phi^0$ and the charged scalars $\phi^\pm$ in the last 
two terms disappear after the symmetry breaking, and turn into the longitudinal components 
of weak gauge bosons. Equation~(\ref{la}) indicates that the Yukawa matrices have been 
diagonalized, but the charged scalar currents persist in the symmetric phase. Note that the 
left-handed charged leptons $\ell^{\prime}_L$, which couple to the right-handed neutrinos 
$\nu_R$ via charged scalar currents, differ from the physical mass eigenstates 
$\ell_L$. The same differentiation applies to $\nu^{\prime}_L$ and $\nu_L$. The 
relations between them are stipulated by the PMNS matrix in Eq.~(\ref{pm}),  
\begin{eqnarray}
\ell_L=U\ell^{\prime}_L,\;\;\;\;
\nu_L=U^\dagger \nu^{\prime}_L;\label{lu1}
\end{eqnarray}
namely, the left-hand sides are the linear combinations of the right-hand sides 
weighted by the PMNS matrix elements. 

The external neutral states $\mathcal{L}_L^{-}\ell_L^{+}$ and $\mathcal{L}_L^{+}\ell_L^{-}$ 
are formed by the charged leptons in the broken phase, so their first emissions are 
composed of real neutral scalars or weak gauge bosons. As the magnitude of the $\mathcal{L}$ 
invariant mass exceeds the restoration scale, internal leptons become massless, and start to 
exchange imaginary neutral scalars and charged scalars. The 
$\mathcal{L}^-_L\ell^+_L$-$\mathcal{L}^+_L\ell^-_L$ mixing must involve 
exchanges of charged scalars $\phi^\pm$ or $W^{\pm}$ bosons in the symmetric phase. The 
Yukawa matrices associated with charged scalar emissions are also diagonal with 
$Y_\ell^{(d)}$ for the $\ell_R\to \nu^{\prime}_L$ transition and $Y_\nu^{(d)}$ for the 
$\nu_R\to \ell^{\prime}_L$ transition. When a right-handed neutrino $\nu_{R}$ emits a 
charged scalar $\phi^+$ and transits to a left-handed charged lepton $\ell_L$, which is one 
of the components of $\ell_L^{\prime}$ according to Eq.~(\ref{lu1}), the emission is 
characterized by a PMNS matrix element. The same PMNS matrix element is 
associated with the $W^+$ boson mission in the $\nu_{L}\to \ell_L$ transition. This 
coincidence is expected, since a charged scalar is equivalent to the longitudinal component of 
a $W$ boson.

The dispersion relation for the mixing amplitude $\Pi(s)\equiv M(s)-i\Gamma(s)/2$ is quoted as 
\cite{Li:2022jxc,Li:2020xrz}
\begin{eqnarray}
M(s)=\frac{1}{2\pi}\int^R ds'\frac{\Gamma(s')}{s-s'}
+\frac{1}{2\pi i}\int_{C_R} ds'\frac{\Pi(s')}{s'-s},\label{dis}
\end{eqnarray}
where $s$ is the mass squared of the heavy charged lepton $\mathcal{L}$, 
and $M(s)$ and $\Gamma(s)/2$ represent the real and imaginary parts of the mixing amplitude, 
respectively. The contour consists of one horizontal path below the branch cut along the 
positive real axis, another horizontal path above the branch cut, and a circle $C_R$ of large 
radius $R$. The invariant mass squared $s$ is situated in a deep Euclidean region inside the
large circle, far away from the branch cut, and its magnitude is above the restoration
scale $\Lambda^2$. We decompose the mixing amplitude into a sum over various intermediate 
neutrino channels,
\begin{eqnarray}
\Pi(s)=\sum_{i,j=1}^3\lambda_{i}\lambda_{j}\Pi_{ij}(s)+\cdots\equiv 
\sum_{i,j=1}^3\lambda_{i}\lambda_{j}\left[M_{ij}(s)-\frac{i}{2}\Gamma_{ij}(s)\right]+\cdots,
\label{ij}
\end{eqnarray} 
where $\lambda_i\equiv U^*_{\mathcal{L} i}U_{\ell i}$ is the product of the PMNS matrix 
elements, and $\cdots$ denote the contributions proportional to higher powers of $\lambda_{i,j}$. 
Because an intermediate state can be identified experimentally in principle, the 
contribution from each channel with the lowest power of $\lambda_i\lambda_j$ satisfies its own 
dispersion relation,
\begin{eqnarray}
M_{ij}(s)=\frac{1}{2\pi}\int^R ds'\frac{\Gamma_{ij}(s')}{s-s'}
+\frac{1}{2\pi i}\int_{C_R} ds'\frac{\Pi_{ij}(s')}{s'-s},\label{d2}
\end{eqnarray}
in which the lower bound of $s'$ for the dispersive integral of $\Gamma_{ij}(s')$ will be 
specified later.


A remark is in order. Here we have assumed that neutrinos are of the Dirac type. If 
neutrinos are of the Majorana type, the analysis will be much more complicated. First, 
the generation of Majorana neutrino masses requires the introduction of symmetry 
breaking mechanism different from the SM one, i.e., of BSM physics. 
New particles, other than the SM Higgs and weak bosons, can thus be 
exchanged in the box diagrams. What BSM mechanism produces the Majorana masses is still 
not clear, such that a definite estimate of related new-physics effects is impossible. Besides, 
the Majorana masses may not vanish together with other SM particle masses above the 
symmetry restoration scale, owing to the aforementioned BSM mechanism. Hence, the Majorana 
masses need to be retained in the loop calculations, which would become extremely lengthy and 
insurmountable. At last, since the Majorana mass terms do not conserve the lepton numbers, 
one can insert mass vertices into internal neutrino lines in a box diagram and convert 
$\mathcal{L}^-$ ($\ell^-$) into $\mathcal{L}^+$ ($\ell^+$). The resultant box diagram then 
depends on an additional product $U_{\mathcal{L}i}^\dagger U_{\mathcal{L} i}$ of the PMNS 
matrix elements, distinct from $\lambda_i=U_{\mathcal{L}i}^\dagger U_{\ell i}$ in 
Eq.~({\ref{ij}}). As a consequence, we have to handle the box-diagram contributions in a 
more intricate decomposition. Nevertheless, the extra parameters from the Majorana phases, 
appearing only in the PMNS matrix elements, do not affect the dispersion relation in 
Eq.~(\ref{d2}). In conclusion, the investigation on the Majorana neutrino case is beyond the 
scope of the present work.

\subsection{Real Contribution}

\begin{figure}
\begin{center}
\includegraphics[scale=0.8]{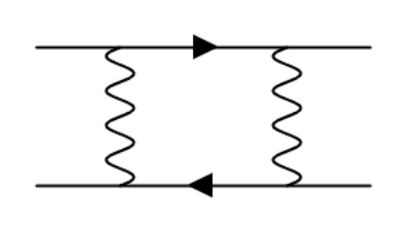}
\caption{\label{fig1}
Box diagram with the two vertical wavy lines representing $W$ bosons.}
\end{center}
\end{figure}

We calculate the real part of the mixing amplitude in the symmetric phase, and collect
the detail in the Appendix. As elucidated before, the external states 
$\mathcal{L}_L^{-}\ell_L^{+}$ and $\mathcal{L}_L^{+}\ell_L^{-}$ exchange only two $W$ bosons 
at one-loop level as depicted by the box diagram in Fig.~\ref{fig1}. The real contribution 
from the box diagrams with the massless internal neutrinos $\nu_{iL}$, $\nu_{jL}$ and two 
massless $W$ bosons is written as
\begin{eqnarray}
M_{ij}^{(1)}(s)=-\frac{1}{16\pi^2}\left(\frac{g}{2\sqrt{2}}\right)^4\frac{4}{s},
\label{m00}
\end{eqnarray}
which is of $O(g^4)$, $g$ being the weak coupling. The above expression is independent of 
intermediate states, so the summation over all intermediate channels vanishes with 
the unitarity condition $\sum_iU^*_{\mathcal{L} i}U_{\ell i}=0$. The absence of 
the mixing in the symmetric phase has been adopted as the input in our previous 
LO analysis \cite{Li:2023ncg}.

\begin{figure}
\begin{center}
\includegraphics[scale=0.8]{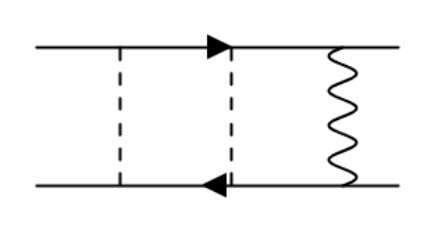}
\caption{\label{fig2}
Two-loop diagram with exchanges of a neutral scalar, a charged scalar and
a $W$ boson in sequence from left to right.}
\end{center}
\end{figure}

At two-loop level, the heavy charged lepton $\mathcal{L}_L^{-}$ can first emit a Higgs 
boson, transiting to a right-handed charged lepton $\mathcal{L}_R^{-}$, which then emits a 
charged scalar $\phi^-$, transiting to a left-handed neutrino $\nu_{iL}$. This $\nu_{iL}$ 
becomes the charged lepton $\ell_L^{-}$ in the external state $\mathcal{L}_L^{+}\ell_L^{-}$ 
finally by emitting a $W^+$ boson. The sequential emissions are illustrated in 
Fig.~\ref{fig2}, in which all the internal particles are massless. The Higgs boson 
vertex carries the Yukawa coupling, i.e., the diagonal element of the Yukawa matrix in 
Eq.~(\ref{la}), $(Y_\ell^{(d)})_{\mathcal{L}\mathcal{L}}/\sqrt{2}=m_{\mathcal{L}}/v$. The 
charged scalar vertex is characterized by the PMNS matrix element $U^*_{\mathcal{L}i}$ and 
the Yukawa coupling $(Y_\ell^{(d)})_{\mathcal{L}\mathcal{L}}=\sqrt{2}m_{\mathcal{L}}/v$ 
according to the Lagrangian in Eq.~(\ref{la}). The $W$ boson vertex contains the PMNS 
matrix element $U_{\ell i}$ and the weak coupling $g$. The product of $U^*_{\mathcal{L}i}$ 
and $U_{\ell i}$ forms the PMNS factor $\lambda_i$. The couplings for the emissions from 
the external charged lepton $\ell_L^{+}$, the internal right-handed charged lepton 
$\ell_R^+$, and the internal left-handed neutrino $\nu_{jL}$ are assigned in a similar way, 
which gives rise to another PMNS factor $\lambda_j$. Th contribution from Fig.~\ref{fig2}, 
unlike Eq.~(\ref{m00}), is counted as $O(g^2)$ and the fourth power in the Yukawa couplings 
for charged leptons.

There are other two-loop diagrams of course. For instance, an additional photon or $Z$ boson
can be exchanged between the charged leptons in Fig.~\ref{fig1},
or a $W$ boson can proceed a hadronic decay, splitting into a quark-antiquark pair. These 
diagrams, being of higher orders in the electroweak coupling, are classified as radiative 
corrections to Fig.~\ref{fig1}. To have them in the real part of the mixing amplitude, the 
corresponding diagrams must be added to the imaginary part for a consistent construction of
the dispersion relation. We will not consider these contributions in the present 
order-of-magnitude investigation. The boson exchanges in the sequence of $W^-$, $\phi^+$ 
and $\phi^0$, opposite to that in Fig.~\ref{fig2}, are permitted. The diagram 
involving a $\phi^0\phi^0\phi^+\phi^-$ four-scalar vertex and a $W$ boson exchange also exists 
at $O(g^2)$. For our purpose, we compute only Fig.~\ref{fig2} in the Appendix, and the result 
reads
\begin{eqnarray}
M_{ij}^{(2)}(s)=-\left(\frac{1}{16\pi^2}\right)^2\left(\frac{g}{2\sqrt{2}}\right)^2
\frac{m_{\mathcal{L}}^2 m_{\ell}^2}{v^4}\frac{4}{s}.\label{m11}
\end{eqnarray}
Equation~(\ref{m11}), vanishing the summation over all intermediate channels, implies that 
the mixing phenomenon in the symmetric phase is still off at two-loop level. 
Note that the lepton mass dependence is brought in via the Yukawa matrix 
$Y^{(d)}_\ell$, though all internal particles are massless in the symmetric phase. 

\begin{figure}
\begin{center}
\includegraphics[scale=0.8]{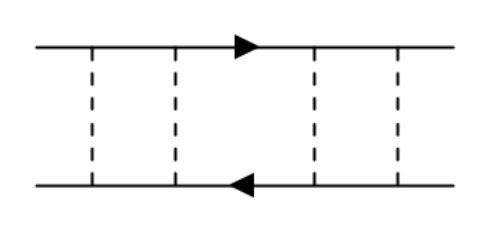}\hspace{1.0cm}
\includegraphics[scale=0.8]{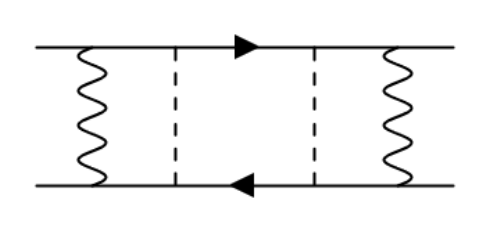}

(a) \hspace{8.0 cm} (b)
\caption{\label{fig3}
(a) Three-loop diagram with exchanges of neutral, charged, charged and neutral scalars
in sequence from left to right. (b) Three-loop diagram with exchanges of a $W$ boson,
two neutral scalars and a $W$ boson in sequence from left to right.}
\end{center}
\end{figure}

We then extend the evaluation of the real part to three-loop level, at which there are even more 
diagrams. The contribution from exchanges of two $W$ bosons and two charged scalars, being 
proportional to the PMNS factor to the fourth power, should be discarded. The diagram with 
exchanges of two photons (or two $Z$ bosons) and two $W$ bosons stands for a higher-order 
correction in the electroweak coupling to Fig.~\ref{fig1}, and is thus neglected. The 
contribution of Fig.~\ref{fig3}(a) with two Higgs bosons and two charged scalars is 
proportional to $O(g^0)$ and the eighth power in the Yukawa couplings for charged leptons,  
different from Eqs.~(\ref{m00}) and (\ref{m11}). At the same order, a Higgs boson can attach 
to $\mathcal{L}_L^{-}$ and $\ell_L^{-}$ and another can attach to $\ell_L^{+}$ 
and $\mathcal{L}_L^{+}$; the two Higgs bosons in the above diagram can cross each other; 
the external leptonic states can interact via two Higgs bosons and a $\phi^+\phi^+\phi^-\phi^-$ 
four-scalar vertex. Figure~\ref{fig3}(b) with two $W$ bosons and two neutral scalars, being 
proportional to $(g^4)$ and the fourth power in the 
Yukawa couplings for neutrinos, also constitutes a higher-order correction to Fig.~\ref{fig1}.
However, it produces the first piece of contributions dependent on intermediate 
neutrino masses, such that the summation over all intermediate channels is finite 
because of $\sum_iU^*_{\mathcal{L} i}U_{\ell i}(Y^{(d)}_\nu)_{ii}^2\not=0$. Therefore, the 
reasoning for the disappearance of the mixing phenomenon in the symmetric phase 
\cite{Li:2023ncg} holds to an extremely high precision in view of the tininess of $Y^{(d)}_\nu$.

We estimate only the $O(g^0)$ contribution from Fig.~\ref{fig3}(a) in the Appendix, obtaining  
\begin{eqnarray}
M_{ij}^{(3)}(s)\approx\frac{1}{64}\left(\frac{1}{16\pi^2}\right)^3
\frac{m_{\mathcal{L}}^4 m_\ell^4}{v^8}\frac{3}{s}.\label{m22}
\end{eqnarray}
The contribution from Fig.~\ref{fig3}(b) is also needed for discriminating the neutrino mass 
orderings in the next section, but its explicit expression is not crucial. We summarize the 
real part of the mixing amplitude in the symmetric phase from Eqs.~(\ref{m00}), (\ref{m11}) 
and (\ref{m22}),
\begin{eqnarray}
M_{ij}(s)\approx-\frac{1}{16\pi^2}\left[4\left(\frac{g}{2\sqrt{2}}\right)^4
+\frac{1}{4\pi^2}\left(\frac{g}{2\sqrt{2}}\right)^2
\frac{m_{\mathcal{L}}^2 m_{\ell}^2}{v^4}
-\frac{3}{64}\left(\frac{1}{16\pi^2}\right)^2\frac{m_{\mathcal{L}}^4 m_\ell^4}{v^8}
+O\left(g^4\frac{m_i^2m_j^2}{v^4}\right)\right]\frac{1}{s},\label{m33}
\end{eqnarray}
where the last term is associated with Fig.~\ref{fig3}(b). Equation~(\ref{m33})
will be inserted into the left-hand side of the dispersion relation in Eq.~(\ref{d2}).

\subsection{Imaginary Contribution}

The neutral leptonic states $\mathcal{L}_L^{-}\ell_L^{+}$ and $\mathcal{L}_L^{+}\ell_L^{-}$ 
mix through the box diagrams with two $W$ boson exchanges in the broken phase. To coordinate 
with the derivation of the real part $M_{ij}(s)$ in the previous subsection, we sum up the 
contributions to the imaginary part $\Gamma_{ij}(s)$ from 
all possible cuts on the internal lines in the box diagrams. This treatment differs from 
that in our earlier work \cite{Li:2023ncg}, where only the two internal neutrino 
lines were cut. It makes sense, since such an intermediate state can be identified 
experimentally in principle, and the corresponding real part vanishes at LO owing to the 
unitarity of the PMNS matrix. The box diagrams induce the $(V-A)(V-A)$ and $(S-P)(S-P)$
structures, which ought to be managed separately. We concentrate on the former, which 
contributes \cite{Cheng,BSS}
\begin{eqnarray}
\Gamma_{ij}(s)&=&\frac{G_F^2}{16\pi}\sum_k'\Gamma_{ij}^k(s),\nonumber\\
\Gamma_{ij}^k(s)&=&\int_{\alpha_k^l}^{\alpha_k^u} d\alpha
\frac{\left(4m_W^4+m_i^2m_j^2\right)F_k(\alpha)-4m_i^2m_j^2m_W^2+2s
\left[\alpha m_i^2+(1-\alpha)m_j^2\right]m_W^2}{(m_W^2-m_i^2)(m_W^2-m_j^2)},
\label{bij}
\end{eqnarray} 
with the notation $\sum'_k\equiv \sum_{k=1}^2-\sum_{k=3}^4$ and the intermediate neutrino 
masses $m_i$ and $m_j$. 
 
The functions $F_k$ take the forms
\begin{eqnarray}
F_1(\alpha)&=&\alpha m_i^2+(1-\alpha)m_j^2-\alpha(1-\alpha)s,\nonumber\\
F_2(\alpha)&=&m_W^2-\alpha(1-\alpha)s,\nonumber\\
F_3(\alpha)&=&\alpha m_W^2+(1-\alpha)m_j^2-\alpha(1-\alpha)s,\nonumber\\
F_4(\alpha)&=&\alpha m_i^2+(1-\alpha)m_W^2-\alpha(1-\alpha)s, \label{ik}
\end{eqnarray}
whose zeros fix the upper and lower bounds of the integration variable $\alpha$, 
\begin{eqnarray}
\alpha_1^{u,l}&=&\frac{1}{2s}\left[s-m_i^2+m_j^2\pm
\sqrt{s^2 - 2s(m_i^2+m_j^2)+(m_i^2-m_j^2)^2}\right],\nonumber\\
\alpha_2^{u,l}&=&\frac{1}{2s}\left[s\pm\sqrt{s^2 - 4sm_W^2}\right],\nonumber\\
\alpha_3^{u,l}&=&\frac{1}{2s}\left[s-m_W^2+m_j^2\pm
\sqrt{s^2 - 2s(m_W^2+m_j^2)+(m_W^2-m_j^2)^2}\right],\nonumber\\
\alpha_4^{u,l}&=&\frac{1}{2s}\left[s-m_i^2+m_W^2\pm
\sqrt{s^2 - 2s(m_i^2+m_W^2)+(m_i^2-m_W^2)^2}\right].\label{ul}
\end{eqnarray}
The indices $k=1$, 2, 3 and 4 correspond to the intermediate states with two real neutrinos, 
two real $W$ bosons, the real neutrino of mass $m_i$ and one real $W$ boson, and the real 
neutrino of mass $m_j$ and one real $W$ boson, respectively. The thresholds for
the dispersive integral in Eq.~(\ref{d2}) are then set to the values $t_k=(m_i+m_j)^2$, 
$4m_W^2$, $(m_i+m_W)^2$ and $(m_j+m_W)^2$, respectively.

It is observed that the linear growth of $\Gamma_{ij}(s)$ at large $s$ \cite{Li:2023ncg} in 
the broken phase cancels among the cuts in Eq.~(\ref{bij}). The cancellation is so strong, 
that $\Gamma_{ij}(s)$ turns out to decrease like $1/s$ asymptotically. It is not hard to 
realize this strong cancelation; the functions $F_k$ in Eq.~(\ref{ik}), as well as the upper 
and lower bounds of $\alpha$ in Eq.~(\ref{ul}), become identical in the $s \to\infty$ limit. 
The real part $M_{ij}(s)$ in Eq.~(\ref{m33}) and the aforementioned decrease of 
$\Gamma_{ij}(s)$ at high $s$ manifest that the mixing amplitude $\Pi_{ij}(s)$ goes down like 
$1/s$. Hence, the contour integral from the big circle diminishes with the radius $R$, and 
the large $s'$ region does not contribute to the dispersive integral in Eq.~(\ref{d2}). This 
fact is also reflected by the $R$ independence of $M_{ij}(s)$ on the left-hand side of 
Eq.~(\ref{d2}). Although the behavior of $\Gamma_{ij}(s)$ in the transition region between 
the symmetry broken and restored phases is unknown, physical insights can be extracted from 
the convergent dispersive integral. First, we remove the interval above the restoration 
scale, $\Lambda^2<s'<R$, motivated by the disappearance of the $R$ dependence. The restoration 
scale, representing an order-of-magnitude concept, can mimic the uncertainty associated with 
the transition region. The integration is then performed with the known integrand $\Gamma_{ij}(s')$ 
in Eq.~(\ref{bij}), which is valid within the interval $t_k<s'<\Lambda^2$. 

Keeping only the terms that survive in the large $\Lambda^2$ limit, we expand the dispersive 
integral as
\begin{eqnarray}
\frac{1}{2\pi}\sum_k'\int_{t_k}^{\Lambda^2} ds'\frac{\Gamma_{ij}^k(s')}{s-s'}
\approx-\frac{G_F^2m_W^4}{16\pi^2}
\left[1 + \frac{m_im_j}{m_W^2}\ln\frac{\Lambda^2}{m_W^2} - \frac{m_i^2m_j^2}{m_W^4}
\left(\ln\frac{m_W^2}{m_im_j} - \frac{1}{4}\right)\right]\frac{2}{s},\label{gG}
\end{eqnarray}
where the approximation $1/(s-s')\approx 1/s$ for $s\gg\Lambda^2$ has been applied,
and those suppressed by higher powers of $m_{i,j}/m_W$ have been dropped for clarity. 
To proceed further, we recast the above expression, by employing the relations
$G_F=\sqrt{2}g^2/(8m_W^2)$ and $m_W=gv/2$, into
\begin{eqnarray}
\frac{1}{2\pi}\sum_k'\int_{t_k}^{\Lambda^2} ds'\frac{\Gamma_{ij}(s')}{s-s'}\approx 
-\frac{1}{16\pi^2}\left[4\left(\frac{g}{2\sqrt{2}}\right)^4 
+ 2\left(\frac{g}{2\sqrt{2}}\right)^2\frac{m_im_j}{v^2}\ln\frac{\Lambda^2}{m_W^2} 
- \frac{m_i^2m_j^2}{v^4}\left(\ln\frac{m_W^2}{m_im_j} - \frac{1}{4}\right)\right]\frac{1}{s}.
\label{gt1}
\end{eqnarray}
It is noticed that powers of $m_W$ have compensated some of the weak coupling, and the 
expansions on both sides of Eq.~(\ref{d2}), i.e., in Eqs.~(\ref{m33}) and (\ref{gt1}), coincide 
perfectly in powers of $g$. In particular, the uncertainty from the transition 
region shows up only in the $O(g^2)$ term. Adding more photons, $Z$ bosons and Higgs bosons 
to the box diagrams just brings in higher-order corrections to Eq.~(\ref{gt1}), which will 
be ignored in the present order-of-magnitude setup.

\section{SOLUTION}

We are ready to acquire information on neutrino masses from Eq.~(\ref{d2}), given the real
piece $M_{ij}(s)$ in the symmetric phase and the imaginary piece $\Gamma_{ij}(s)$ in the 
broken phase. Our framework is appropriate for the mixing of heavy-light systems, 
like $\mu e$ and $\tau e$. This claim is justified by $\mu$-$\tau$ refection symmetry 
observed in our previous work \cite{Li:2023ncg}; the ratio 
$U^*_{\mu 1}U_{e1}/(U^*_{\mu 2}U_{e2})$ of the PMNS matrix elements for the $\mu e$ mixing and 
$U^*_{\tau 1}U_{e1}/(U^*_{\tau 2}U_{e2})$ for the $\tau e$ mixing are roughly equal, except 
the opposite signs of their imaginary parts. This prediction for heavy-light systems agrees 
well with the outcomes of global fits \cite{PDG}. However, the ratio 
$U^*_{\tau 1}U_{\mu 1}/(U^*_{\tau 2}U_{\mu 2})$ for the $\tau \mu$ mixing does not follow the 
same pattern, suggesting that $\tau \mu$ should not be regarded as a heavy-light system. The 
distinction is natural in view of the charged lepton masses; a muon is about 200 times heavier 
than an electron, while a $\tau$ lepton is only 17 times heavier than a muon.

We have shown in Sec.~IIA that $M_{ij}(s)$ in the symmetric phase is proportional to the 
masses of the external charged leptons. To probe how small a typical neutrino mass can be, 
we consider the $\mu e$ mixing, and set $m_{i,j}$ in Eq.~(\ref{gt1}) 
to a common mass scale $m_\nu$. As seen shortly, the value of $m_\nu$ determined by the 
dispersion relation is of order eV, so it is not necessary to differentiate which generation 
of neutrinos this typical scale is assigned to; an eV scale is much higher than the 
measured mass differences 
$\Delta m^2_{21} \equiv m^2_{2}-m^2_{1}= (7.55^{+0.20}_{-0.16})\times 10^{-5}$ eV$^2$ 
and $\Delta m^2_{32}\equiv m^2_{3}-m^2_{2}=(2.424\pm 0.03)\times 10^{-3}$ eV$^2$ 
from the global fit \cite{deSalas:2017kay}. We create a solution to the dispersion relation 
by equating the terms in Eqs.~(\ref{m33}) and ~(\ref{gt1}) with the same powers of the weak 
coupling $g$, for $g$ can vary freely in a mathematical viewpoint. 
It is immediately found that the first, i.e., $O(g^4)$, terms in Eqs.~(\ref{m33}) and (\ref{gt1}) 
are the same. We emphasize that this equality is insensitive to the behavior of $\Gamma_{ij}(s)$ 
in the transition region, namely, independent of the symmetry restoration scale.

The equality of the third, i.e., $O(g^0)$, terms in Eqs.~(\ref{m33}) and (\ref{gt1}) leads to 
Eq.~(\ref{mn3}), where $m_{\mathcal{L}}$, $m_\ell$ and $m_{i,j}$ have been replaced by the 
muon mass $m_\mu$, the electron mass $m_e$, and the typical neutrino mass $m_\nu$, respectively. 
The constant $1/4$ is negligible compared with the logarithmic term $\ln(m_W^2/m_\nu^2)$. 
Equation~(\ref{mn3}), arising from the convergent part of the dispersive integral, does not 
depend on the restoration scale either. 
It then designates a small mass scale for neutrinos unambiguously, 
\begin{eqnarray}
m_\nu\approx 3\;{\rm eV},\label{2eV}
\end{eqnarray}
with the inputs $m_W\approx 80$ GeV, $m_\mu\approx 0.1$ GeV, $m_e\approx 0.5$ MeV and 
$v\approx 250$ GeV \cite{PDG}. Note that such a low mass scale $m_\nu\sim O(1)$ eV is demanded 
by the internal consistency the SM dynamics, whose origin is quite different from the seesaw 
mechanism \cite{see} as stated in the Introduction. Our order-of-magnitude result is 
not far beyond the upper bound on the neutrino mass $m_\nu<0.9$ eV at 90\% CL measured by the 
KATRIN Collaboration \cite{Nature}. We point out that their measurement is independent of 
cosmological models and does not rely on assumptions whether a neutrino is a Dirac or 
Majorana particle.

The Planck Collaboration reported the sum of neutrino masses $\sum m_\nu  < 0.12$ eV at 
95\% CL by analyzing the Cosmic Microwave Background (CMB) and Baryon Acoustic Oscillations 
(BAO) data within the standard cosmological model $\Lambda$CDM  \cite{Planck:2018vyg}.
Nevertheless, the authors in \cite{Alvey:2021sji} showed that neutrino masses as large as 
$\sum m_\nu \sim 3$ eV, i.e., of eV order, are perfectly consistent with the CMB and BAO data,
if the neutrino number density is allowed to deviate from the assumed Fermi-Dirac distribution. 
Another model dependence from the inhomogeneous reionization effect was addressed in 
\cite{Long:2022dil}. The recent BAO measurements 
from the Dark Energy Spectroscopic Instrument (DESI) have led to even tighter upper 
bounds on the neutrino mass sum, $\sum m_\nu<0.072$ (0.113) eV at 95\% CL for 
a $\sum m_\nu>0$ ($\sum m_\nu>0.059$) eV prior \cite{DESI:2024mwx},
exhibiting potential tension with the laboratory oscillation data. Revised 
or extended models have been proposed to reconcile the tension in the literature, which 
disclose the subtlety in the cosmological constraints. For this reason, we
prefer not to confront our findings with the cosmological constraints.


Next we equate the second, i.e., $O(g^2)$, terms in Eqs.~(\ref{m33}) and (\ref{gt1}),
reaching
\begin{eqnarray}
m_\nu^2\ln\frac{\Lambda}{m_W}\approx \frac{1}{16\pi^2}
\frac{m_\mu^2 m_e^2}{v^2},\label{22}
\end{eqnarray}
whose combination with Eq.~(\ref{mn3}) yields Eq.~(\ref{mn32}). It is stressed that 
Eq.~(\ref{mn32}) is independent of the masses $m_{\mathcal{L}}$ and $m_\ell$ of the external 
charged leptons. Because of the uncertain behavior of $\Gamma_{ij}(s)$ in the transition region, 
we have regarded the coefficient on the right-hand side of Eq.~(\ref{mn32}) as being of $O(1)$. 
Unfortunately, the value of the restoration scale $\Lambda$ inferred from Eq.~(\ref{mn32}) is 
sensitive to the $O(1)$ coefficient, so no definite result can be obtained (for a very crude 
guesstimate, Eq.~(\ref{2eV}) corresponds to $\Lambda\gtrsim O(100)$ TeV via Eq.~(\ref{mn32})). 
Instead, we highlight the essential indication that the large electroweak symmetry restoration, 
i.e., new physics scale is tied to the tiny neutrino mass. A precise prediction for the 
new physics scale can be achieved by exploring $\Gamma_{ij}(s)$ in the transition region and by 
calculating exact multi-loop contributions to $M_{ij}(s)$ in the symmetric phase.

At last, we sum both Eq.~(\ref{m33}) and (\ref{gt1}) over all intermediate channels, and
check whether the constraint on the PMNS matrix elements and the neutrino mass 
orderings in our previous LO study \cite{Li:2023ncg} is maintained. As explained before, the 
contribution to the real piece $M_{ij}(s)$ in the symmetric phase, which depends on the 
intermediate neutrino masses and survives the unitarity of the PMNS 
matrix, starts at $O(m_i^2m_j^2/v^4)$. The summation for 
Eq.~(\ref{gt1}) is dominated by the second term of $O(m_im_j/v^2)$. The elimination of 
$\lambda_3=-\lambda_1-\lambda_2$ by means of the unitarity condition generates
\begin{eqnarray}
& &\ln\frac{\Lambda^2}{m_W^2}[\lambda_1(m_3-m_1)+ \lambda_2(m_3- m_2)]^2=\sum_{i,j}
O\left(\frac{m_i^2m_j^2}{v^2}\right)\approx 0,\label{mimj}
\end{eqnarray}
which gives rise to Eq.~(\ref{mn33}) trivially irrespective of the restoration scale 
$\Lambda$; the $O(m_im_j)$ coefficient of $\ln(\Lambda^2/m_W^2)$ must diminish to match the 
right-hand side of $O(m_i^2m_j^2/v^2)$. 

To facilitate the discussion, we adopt the PMNS matrix in the Chau-Keung 
parametrization \cite{CK84},
\begin{equation}           
U_{\rm PMNS}=\left( \begin{array}{ccc}
    c_{12}c_{13} & s_{12}c_{13} & s_{13}e^{-i\delta}\\
    -s_{12}c_{23}-c_{12}s_{23}s_{13}e^{i\delta} & c_{12}c_{23}-s_{12}s_{23}s_{13}e^{i\delta}&
    s_{23}c_{13}\\
    s_{12}s_{23}-c_{12}c_{23}s_{13}e^{i\delta} & -c_{12}s_{23}-s_{12}c_{23}s_{13}e^{i\delta} &
    c_{23}c_{13}
   \end{array} \right)\;,\label{CKMb}
\end{equation} 
with the notations $s_{12}\equiv\sin\theta_{12}$, $c_{12}\equiv\cos\theta_{12}$, etc. for
the mixing angles, and the $CP$ phase $\delta$. The products of the PMNS matrix elements then 
read 
\begin{eqnarray}
\lambda_1=-c_{12}c_{13}\left(s_{12}c_{23}+c_{12}s_{23}s_{13}e^{-i\delta}\right),\;\;\;\; 
\lambda_2=s_{12}c_{13}\left(c_{12}c_{23}-s_{12}s_{23}s_{13}e^{-i\delta}\right). 
\label{a12}
\end{eqnarray}
Defining $r=\lambda_1/\lambda_2$, we require the cancellation between the real parts of 
$r(m_3-m_1)$ and $(m_3- m_2)$ according to Eq.~(\ref{mn33}),
\begin{eqnarray}
{\rm Re}r(m_3-m_1)+(m_3- m_2)\approx 0,\label{mm31}
\end{eqnarray}
and the small imaginary part of $r(m_3-m_1)$, i.e., a small $s_\delta\equiv\sin\delta$. 

It is easy to verify 
that the mixing angles $\theta_{12}= (34.5^{+1.2}_{-1.0})^{\circ}$, 
$\theta_{13}=(8.45^{+0.16}_{-0.14})^{\circ}$ and $\theta_{23}=(47.7^{+1.2}_{-1.7})^{\circ}$, 
and the $CP$ phase $\delta=(218^{+38}_{-27})^{\circ}$ from the global fit 
\cite{deSalas:2017kay} in the NO scenario meet better the above requirements, compared to 
$\theta_{12}=(34.5^{+1.2}_{-1.0})^{\circ}$, $\theta_{13}=(8.53^{+0.14}_{-0.15})^{\circ}$, 
$\theta_{23}=(47.9^{+1.0}_{-1.7})^{\circ}$ and $\delta=(281^{+23}_{-27})^{\circ}$ in the IO 
one \cite{deSalas:2017kay}. We set $\theta_{12}$, $\theta_{13}$ and $\theta_{23}$ to their
central values viewing their small errors, and vary $\delta$ within its 1-$\sigma$ range.
Equation~(\ref{mm31}) can be satisfied by $c_\delta<0$ (${\rm Re}r>-1$) and $m_3-m_1>0$ 
for the NO, and by $c_\delta>0$ (${\rm Re}r<-1$) and $m_3-m_1<0$ for the IO. 
The whole range of $\delta$ for the NO is acceptable, for the resultant ${\rm Re}r$ takes 
values between $-0.9$ and $-0.7$, while $\delta$ in the IO case is restricted to 
$(270^\circ, 304^\circ)$. We then have $s_\delta\approx -(0.62^{+0.35}_{-0.43})$ from the 
NO, whose magnitude could be much lower than that of $s_\delta\in (-1,-0.83)$ from the IO.
The latter narrow range is attributed to that $\delta$ for the IO is about $3\pi/2$, so the 
amount of errors similar to the NO case causes minor variation of $s_\delta$. Our solution 
favors negative $c_\delta$ and smaller $|s_\delta|$, close to that derived in a 
seesaw model with right-handed neutrino masses being dynamically generated by strong 
horizontal gauge interactions \cite{Chung:2023rie}.



\section{CONCLUSION}

We have proposed a viable explanation on the extremely small neutrino masses, which comes 
as a straightforward extension of our continuous efforts in understanding the SM flavor 
structure. Dispersion relations, which analytical physical observables must obey, 
bind interactions at various scales, and enforce stringent constraints on 
the SM parameters. The assumption for the existence of an electroweak symmetric phase
grants such a bond between the dynamics above and below the symmetry restoration scale. 
We have taken into account the contributions to the mixing of neutral leptonic states up to 
three loops in the symmetric phase, which arise from the exchanges of neutral and charged 
scalars and $W$ bosons. The standard box diagrams responsible for the mixing in the 
symmetry broken phase involve the neutrino masses and the PMNS matrix elements. It was then 
demonstrated that the dispersion relation for the $\mu^-e^+$-$\mu^+e^-$ mixing 
demands a typical neutrino mass scale $m_\nu$ at eV order in the SM unambiguously. It is 
much lower than a lepton mass because of the huge suppression factors $m_\ell/v$ formed by 
the lepton mass and the electroweak scale, and from the measure of the three-loop integral. 
At two-loop level, the dispersion relation hints that the large electroweak symmetry 
restoration, i.e., new physics scale is linked to the small neutrino mass. 

The computation presented in the Appendix confirmed that the disappearance of the mixing 
phenomenon in the symmetric phase, argued in our previous publication based on the unitarity 
of the PMNS matrix, holds to high precision, up to corrections of $O(m_\nu^4/v^4)$. In 
other words, the first contribution which survives the summation over all 
intermediate channels begins at three-loop level. It is thus expected that the dispersive 
constraint on the neutrino masses and the PMNS matrix elements still prefers the NO over the 
IO under the loop corrections. The pivot for discriminating the neutrino mass orderings 
is played by the $CP$ phase, which got distinct results in the global fits for the NO and 
the IO. Since it is challenging to settle the neutrino mass spectrum experimentally, our 
theoretical attempt should be valuable. The above summaries the solution to the dispersion 
relation for the mixing of heavy-light leptonic systems. We admit that our attempt is 
preliminary, accurate only to orders of magnitude, and needs to be refined in the future. 
More definite predictions for the neutrino masses and for the new physics scale 
can be made by exploring the imaginary piece $\Gamma_{ij}(s)$ in the transition region and 
by assessing complete multi-loop contributions to the real piece $M_{ij}(s)$ in the symmetric 
phase.

We emphasize that the current mechanism for generating a small neutrino mass cannot be applied 
to the quark sector, simply due to strong interaction among quarks. The strong coupling 
$g_s$ is much larger than most of the quark Yukawa couplings, even greater than 
the top Yukawa coupling (for the magnitudes of various couplings with 
renormalization-group effects in the SM, refer to \cite{Buttazzo:2013uya}), 
such that the loop corrections discussed in this work are overwhelmed by 
QCD ones. For example, the two neutral scalars in Fig.~\ref{fig3}(a) for the mixing between 
the neutral quark states $c\bar u$ and $\bar c u$ can be replaced by two gluons. The resultant 
contribution then dominates over the former one by a factor $g_s^4 v^4/(m_c^2m_u^2)\gg 1$ with 
the charm (up) quark mass $m_c$ ($m_u)$. It is obvious that Eq.~(\ref{mn3}) is not justified. 
In this sense, the origin of the tiny neutrino 
mass traces back to the fact that a neutrino participates only weak and scalar interactions. 
Combining the present and previous observations, we conjecture that the mass hierarchy inherent 
in the SM, from the neutrino masses up to the electroweak scale, and the distinct quark and 
lepton mixing patterns, may be accommodated by means of the internal consistency of the SM 
dynamics. This understanding also points to potential new physics scenarios above the electroweak 
symmetry restoration scale, like the composite Higgs model, which will be investigated in 
forthcoming papers.

\section*{Acknowledgement}

We thank Y. Chung, X.G. He, Y.B. Hsiung, Y. Liao, J.L. Liu and M.R. Wu for illuminating 
discussions. This work was supported in part by National Science and Technology Council of 
the Republic of China under Grant No. MOST-110-2112-M-001-026-MY3.

\appendix
\section{MIXING AMPLITUDE UP TO THREE LOOPS IN SYMMETRIC PHASE}

\subsection{One Loop}

The box-diagram contribution, including both the real and imaginary parts, has been derived 
in \cite{BSS}. We set the intermediate neutrino masses to zero, and then take the $m_W\to 0$ 
limit to get the leading real contribution in the symmetric phase shown in Eq.~(\ref{m00}). 
The dispersion relation at $O(g^4)$ is then verified by 
Eqs.~(\ref{m33}) and (\ref{gt1}). To reduce the load for two-loop and three-loop calculations, 
we propose a simplified scheme below. The amplitude from the box diagram in Fig.~\ref{fig1} 
for the $\mathcal{L}^-\ell^+$-$\mathcal{L}^+\ell^-$ mixing with 
massless intermediate particles is written as
\begin{eqnarray}
-i{\cal M}=\frac{1}{2}\left(\frac{-ig}{2\sqrt{2}}\right)^4\int \frac{d^4l}{(2\pi)^4}
\frac{{\bar u}_\ell\gamma_\nu(1-\gamma_5)i(\not p-\not l)\gamma_\mu(1-\gamma_5) u_{\mathcal{L}}
{\bar v}_\ell\gamma_{\mu'}(1-\gamma_5)(-i)\not l\gamma_{\nu'}(1-\gamma_5) v_{\mathcal{L}}
(-i)^2g^{\nu\nu'}g^{\mu\mu'}}{(l^2+i\epsilon)^2[(p-l)^2+i\epsilon]^2}.\label{a1}
\end{eqnarray}
The PMNS matrix elements have been suppressed, the spin factor $1/2$ comes from the 
definition of the neutral leptonic state, the coupling $-ig/(2\sqrt{2})$ is associated with 
a weak vertex, and the tensors $g^{\nu\nu'}$ and $g^{\mu\mu'}$ are from the massless $W$ 
boson propagators. Strictly speaking, the aforementioned $m_W\to 0$ limit urges that we 
start with a massive $m_W$ boson propagator, 
\begin{eqnarray}
-i\frac{g^{\mu\mu'}-l^\mu l^{\mu'}/m_W^2}{l^2-m_W^2},
\end{eqnarray}
involving an additional longitudinal polarization. Here we employ a massless $W$ boson propagator 
directly for simplification.

The Fierz identity is inserted into Eq.~(\ref{a1}) to factorize the lepton currents,
\begin{eqnarray}
{\cal M}&=&-\frac{1}{4}\left(\frac{g}{2\sqrt{2}}\right)^4
{\bar u}_\ell\gamma^\alpha(1-\gamma_5) u_{\mathcal{L}}
{\bar v}_\ell\gamma^\beta(1-\gamma_5)v_{\mathcal{L}}i\int \frac{d^4l}{(2\pi)^4} 
\frac{{\rm tr}[\gamma_\mu\gamma_\alpha\gamma^{\mu}\not l\gamma^\nu
\gamma_\beta\gamma_\nu(\not p-\not l)
(1+\gamma_5)]}{(l^2+i\epsilon)^2[(p-l)^2+i\epsilon]^2}.
\end{eqnarray}
If we work on the $(S-P)(S-P)$ structure, the term $(1-\gamma_5)\otimes(1-\gamma_5)$
in the Fierz identity will need to be inserted too, and the derivation will be more cumbersome. 
To extract the $(V-A)(V-A)$ current, we delete the current product
${\bar u}_\ell\gamma^\alpha(1-\gamma_5) u_{\mathcal{L}}
{\bar v}_\ell\gamma^\beta(1-\gamma_5)v_{\mathcal{L}}$, and
contract the remaining piece by the projector $g^{\alpha\beta}/4$. The projector should 
be $(g^{\alpha\beta}-p^\alpha p^\beta/p^2)/3$ to avoid the $(S-P)(S-P)$ component. However, we 
have confirmed that $g^{\alpha\beta}/4$ reproduces Eq.~(\ref{m00}) from Eq.~(\ref{a1}) with 
massless $W$ boson propagators. This simpler scheme will be applied to the evaluations of 
two-loop and three-loop diagrams for an order-of-magnitude estimate. We acquire from 
Fig.~\ref{fig1}
\begin{eqnarray}
{\cal M}&=&2\left(\frac{g}{2\sqrt{2}}\right)^4i\int \frac{d^4l}{(2\pi)^4} 
\frac{l\cdot(p-l)}{(l^2+i\epsilon)^2[(p-l)^2+i\epsilon]^2}
=\frac{2}{16\pi^2}\left(\frac{g}{2\sqrt{2}}\right)^4\frac{1}{s},\label{one}
\end{eqnarray}
with $s=p^2$, which is assumed to be in the Euclidean region. 

We have quoted the imaginary contribution of the box diagrams in the broken phase 
from \cite{BSS} as shown in Sec.~II. It is reminded that the analysis in \cite{BSS} 
was conducted for the mixing of neutral quark states. A quark possesses degrees of freedom from 
colors, and the box diagrams with vertical $W$ boson lines (in Fig.~\ref{fig1}) and with 
horizontal $W$ boson lines have different color factors, 1 for the former and 3 for the latter. 
Since a color factor can vary arbitrarily in a mathematical viewpoint, each box diagram 
respects its own dispersion relation. We will take advantage of this fact for the leptonic state 
mixing, combining the above two box diagrams in the way the same as in the quark case, although 
a lepton has no color charge. The contribution from the second box diagram with horizontal 
$W$ boson lines turns out to be identical to Eq.~(\ref{one}) but with an opposite sign. The 
combination of the two box diagrams then yields Eq.~(\ref{m00}). We do not include the diagram 
with the two $W$ bosons in the second diagram crossing each other, because it also obeys a separate 
dispersion relation with different branch cuts; the two intermediate neutrinos of unequal 
masses cannot be on shell simultaneously in this case.

\subsection{Two Loops}

The two-loop diagram in Fig.~\ref{fig2} contributes in the symmetric phase
\begin{eqnarray}
-i{\cal M}&=&\frac{1}{2}\left(-i\frac{m_{\mathcal{L}}}{v}\right)
\left(-i\frac{\sqrt{2}m_{\mathcal{L}}}{v}\right)
\left(-i\frac{m_\ell}{v}\right)\left(-i\frac{\sqrt{2}m_\ell}{v}\right)
\left(\frac{-ig}{2\sqrt{2}}\right)^2
\int \frac{d^4l_1}{(2\pi)^4}\int \frac{d^4l_2}{(2\pi)^4}\nonumber\\
& &\times
\frac{{\bar u}_\ell\gamma_\mu(1-\gamma_5)i(\not p-\not l_2)i(\not p-\not l_1)u_{\mathcal{L}}
{\bar v}_\ell(-i)\not l_1(-i)\not l_2\gamma_{\nu}(1-\gamma_5) v_{\mathcal{L}}}
{(p-l_2)^2(p-l_1)^2l_1^2l_2^2}\frac{i}{l_1^2}\frac{i}{(l_2-l_1)^2}\frac{-ig^{\mu\nu}}{(p-l_2)^2},
\label{m2}
\end{eqnarray}
where the Yukawa couplings for the neutral and charged scalar emissions have been specified 
in Sec.~II. Inserting the Fierz identity to factorize the lepton currents and 
contracting the remaining part by the projector $g^{\alpha\beta}/4$, we have
\begin{eqnarray}
{\cal M}&=&-\frac{m_{\mathcal{L}}^2 m_\ell^2}{v^4}\left(\frac{g}{2\sqrt{2}}\right)^2
\int \frac{d^4l_1}{(2\pi)^4}\int \frac{d^4l_2}{(2\pi)^4}
\frac{l_1\cdot l_2(p-l_2)\cdot(p-l_1)}
{((p-l_2)^2)^2(p-l_1)^2(l_1^2)^2l_2^2(l_2-l_1)^2}.\label{m01}
\end{eqnarray}
We first work out the integration over $l_1$ under the Feynman parametrization,
\begin{eqnarray}
\int\frac{d^4l_1}{(2\pi)^4}\frac{l_1\cdot l_2(p-l_2)\cdot(p-l_1)}
{(l_1^2)^2(p-l_1)^2(l_2-l_1)^2}
=6\int_0^1 dv\int_0^{1-v}du(1-u-v)\int\frac{d^4l_1}{(2\pi)^4}
\frac{l_1\cdot l_2(p-l_2)\cdot(p-l_1)}
{[l_1^2-2up\cdot l_1+us-2vl_2\cdot l_1+vl_2^2]^4}.
\end{eqnarray}
The variable change $l_1\to l_1+up+vl_2$ and the Wick rotation lead the above integral to
\begin{eqnarray}
& &\frac{i\pi^2}{(2\pi)^4}\int_0^1 dv\int_0^{1-v}du(1-u-v)\left\{
\frac{(up+vl_2)\cdot l_2(p-l_2)\cdot[(1-u)p-vl_2]}
{[v(1-v)l_2^2-2uvp\cdot l_2+u(1-u)s]^2}\right.\nonumber\\
& &\hspace{5.0cm}\left.
-\frac{(p-l_2)\cdot l_2}{2[v(1-v)l_2^2-2uvp\cdot l_2+u(1-u)s]}\right\}.\label{98}
\end{eqnarray}

We then turn to the integration over $l_2$. The first piece in Eq.~(\ref{98}) gives,
under the Feynman parametrization,
\begin{eqnarray}
I_1&=&\int \frac{d^4l_2}{(2\pi)^4}
\frac{(up+vl_2)\cdot l_2(p-l_2)\cdot[(1-u)p-vl_2]}
{l_2^2((p-l_2)^2)^2[v(1-v)l_2^2-2uvp\cdot l_2+u(1-u)s]^2}\nonumber\\
&=&\int_0^1 dw\int_0^{1-w}dt\frac{24tw}{v^2(1-v)^2}\int \frac{d^4l_2}{(2\pi)^4}
\frac{(up+vl_2)\cdot l_2(p-l_2)\cdot[(1-u)p-vl_2]}
{\{l_2^2-2tp\cdot l_2-2uwp\cdot l_2/(1-v)+ts+u(1-u)ws/[v(1-v)]\}^5}.\label{a9}
\end{eqnarray}
The variable change 
\begin{eqnarray}
l_2\to l_2+\left(t+\frac{uw}{1-v}\right)p\label{100}
\end{eqnarray}
brings Eq.~(\ref{a9}) into
\begin{eqnarray}
I_1
&=&\frac{i\pi^2}{(2\pi)^4}\int_0^1 dw\int_0^{1-w}dt\frac{2tw}{(1-v)^2}
\frac{A (B + D) + (B-C)D+ 3D^2}{D^3},
\end{eqnarray}
with the functions
\begin{eqnarray}
A&=&\left(\frac{u}{v}+t+\frac{uw}{1-v}\right)\left(t+\frac{uw}{1-v}\right)s\nonumber\\
B&=&\left(1-t-\frac{uw}{1-v}\right)\left(\frac{1-u}{v}-t-\frac{uw}{1-v}\right)s\nonumber\\
C&=&\left(\frac{u}{v}+2t+\frac{2uw}{1-v}\right)
\left(1+\frac{1-u}{v}-2t-\frac{2uw}{1-v}\right)\frac{s}{4}\nonumber\\
D&=&\left[t+\frac{u(1-u)w}{v(1-v)}\right]s-\left(t+\frac{uw}{1-v}\right)^2s.
\end{eqnarray}
We calculate the second term in Eq.~(\ref{98}) in a similar way,
\begin{eqnarray}
I_2&=&\frac{1}{2}\int \frac{d^4l_2}{(2\pi)^4}
\frac{(l_2-p)\cdot l_2}{l_2^2((p-l_2)^2)^2[v(1-v)l_2^2-2uvp\cdot l_2+u(1-u)s]}\nonumber\\
&=&\int_0^1 dw\int_0^{1-w}dt\frac{6t}{2v(1-v)}\int \frac{d^4l_2}{(2\pi)^4}\frac{(l_2-p)\cdot l_2}
{\{l_2^2-2tp\cdot l_2-2uwp\cdot l_2/(1-v)+ts+u(1-u)ws/[v(1-v)]\}^4}.
\end{eqnarray}
The same variable change in Eq.~(\ref{100}) generates 
\begin{eqnarray}
I_2
=\frac{i\pi^2}{(2\pi)^4}\int_0^1 dw\int_0^{1-w}dt\frac{t}{2v(1-v)}\frac{2D - E}{D^2},
\end{eqnarray}
with the function
\begin{eqnarray}
E=\left(1-t-\frac{uw}{1-v}\right)\left(t+\frac{uw}{1-v}\right)s.
\end{eqnarray}

At last, the contribution from Fig.~\ref{fig2} is given by 
\begin{eqnarray}
{\cal M}&=&-\frac{m_{\mathcal{L}}^2 m_\ell^2}{v^4}\left(\frac{g}{2\sqrt{2}}\right)^2
\frac{i\pi^2}{(2\pi)^4}\int_0^1 dv\int_0^{1-v}du(1-u-v)(I_1+I_2)\nonumber\\
&=&\frac{m_{\mathcal{L}}^2 m_\ell^2}{v^4}\left(\frac{g}{2\sqrt{2}}\right)^2
\left(\frac{1}{16\pi^2}\right)^2\frac{2}{s}.
\end{eqnarray}
The result from the diagram with horizontal boson lines has the same 
expression but with an opposite sign. Following the special combination of the two 
diagrams, we arrive at Eq.~(\ref{m11}).

\subsection{Three Loops}

The three-loop diagram with four vertical scalar lines in Fig.~\ref{fig3}(a) contains the 
loop integral
\begin{eqnarray}
-i{\cal M}&=&\frac{1}{2}\left(-i\frac{m_{\mathcal{L}}}{v}\right)^2
\left(-i\frac{\sqrt{2}m_{\mathcal{L}}}{v}\right)^2
\left(-i\frac{m_\ell}{v}\right)^2\left(-i\frac{\sqrt{2}m_\ell}{v}\right)^2
\int \frac{d^4l_1}{(2\pi)^4}\int \frac{d^4l_2}{(2\pi)^4}\int \frac{d^4l_3}{(2\pi)^4}\nonumber\\
& &\times\frac{{\bar u}_\ell i\not l_3i\not l_2i(\not p-\not l_1)u_{\mathcal{L}}
{\bar v}_\ell(-i)\not l_1(-i)(\not p-\not l_2)(-i)(\not p-\not l_3) v_{\mathcal{L}}}
{(p-l_1)^2(p-l_2)^2(p-l_3)^2 l_1^2l_2^2l_3^2}
\frac{i}{l_1^2}\frac{i}{(p-l_1-l_2)^2}\frac{i}{(l_2-l_3)^2}\frac{i}{l_3^2}.\label{3b1}
\end{eqnarray}
We have arranged the loop momentum flows to facilitate the identification of the leading
regions for the integral. It is straightforward to tell that finite $l_1$ gives an important 
contribution, for both the infrared and ultraviolet regions are suppressed. Small $l_2$ 
dominates, in which the integration over $l_3$ develops an infrared logarithmic enhancement. 
We thus neglect $l_2$ in the factor $p-l_1-l_2$, such that the integration over $l_1$ is 
simplified to the one for the box diagram in Fig.~\ref{fig1}, and retain only the terms up to 
those linear in $l_2$ in the integral over $l_3$. An order-of-magnitude estimate serves our 
purpose, so the above approximation is appropriate, without which the handling of three-loop 
diagrams would be extremely tedious. The reasoning then motivates the factorization
of Eq.~(\ref{3b1}) into three pieces by inserting the Fierz identity,
\begin{eqnarray}
{\cal M}&=&\frac{2i}{8^4}\frac{m_{\mathcal{L}}^4 m_\ell^4}{v^8}
{\bar u}_\ell\gamma^\alpha(1-\gamma_5) u_{\mathcal{L}}
{\bar v}_\ell\gamma^\beta(1-\gamma_5)v_{\mathcal{L}}\nonumber\\
& &\times\int \frac{d^4l_2}{(2\pi)^4}\frac{{\rm tr}[ (\not p-\not l_2)
\gamma^\lambda(1-\gamma_5)\not l_2\gamma^\sigma(1-\gamma_5)]}{(p-l_2)^2l_2^2} \nonumber\\
& &\times\int \frac{d^4l_1}{(2\pi)^4}
\frac{{\rm tr}[\not l_1(1-\gamma_5)\gamma_\sigma
(\not p-\not l_1)(1-\gamma_5)\gamma_\alpha]}{(p-l_1)^2(p-l_1-l_2)^2(l_1^2)^2}\nonumber\\
& &\times\int\frac{d^4l_3}{(2\pi)^4}\frac{{\rm tr}[(\not p-\not l_3)
(1-\gamma_5)\gamma_\beta\not l_3(1-\gamma_5)\gamma_\lambda]}
{(p-l_3)^2(l_3^2)^2(l_2-l_3)^2}.
\end{eqnarray}

The integral over $l_1$ is given by
\begin{eqnarray}
J_1&=&\int \frac{d^4l_1}{(2\pi)^4}\frac{{\rm tr}[\not l_1(1-\gamma_5)\gamma_\sigma
(\not p-\not l_1)(1-\gamma_5)\gamma_\alpha]}{(p-l_1)^2(p-l_1-l_2)^2(l_1^2)^2}
\approx 2\int \frac{d^4l_1}{(2\pi)^4}\frac{{\rm tr}[\not l_1\gamma_\sigma
(\not p-\not l_1)\gamma_\alpha(1+\gamma_5)]}{((p-l_1)^2)^2(l_1^2)^2},\label{j1}
\end{eqnarray}
in the small $l_2$ region. The matrix $\gamma_5$ does not contribute, because the vectors 
$p$ and $l_1$ must appear in even powers after the Feynman parametrization. 
Equation~(\ref{j1}) becomes
\begin{eqnarray}
J_1&\approx&24\int_0^1 du u(1-u)\int \frac{d^4l_1}{(2\pi)^4}\frac{4u(1-u)p_\alpha p_\sigma
-g_{\alpha\sigma}[l^2+2u(1-u)s]}{[l_1^2-u(1-u)s]^4}
=\frac{i}{\pi^2}\frac{p_\alpha p_\sigma}{s^2},
\end{eqnarray} 
where the portion proportional to $g_{\alpha\sigma}$ turns out to vanish. 

The third integral is computed as
\begin{eqnarray}
J_3&=&\int\frac{d^4l_3}{(2\pi)^4}\frac{{\rm tr}[(\not p-\not l_3)
(1-\gamma_5)\gamma_\beta\not l_3(1-\gamma_5)\gamma_\lambda]}
{(p-l_3)^2(l_3^2)^2(l_2-l_3)^2}\\
&=&12\int_0^1 du\int_0^{1-u}dv(1-u-v)i\int \frac{d^4l_3}{(2\pi)^4}\nonumber\\
& &\times\frac{2(up+vl_2)_\lambda[(1-u)p-vl_2]_\beta+2(up+vl_2)_\beta[(1-u)p-vl_2]_\lambda
-g_{\lambda\beta}\{l_3^2+2(up+vl_2)\cdot[(1-u)p-vl_2]\}}
{[l_3^2-u(1-u)s-v(1-v)l_2^2+2uvp\cdot l_2]^4},\nonumber
\end{eqnarray}
to which the Feynman parametrization and the variable change $l_3\to l_3+up+vl_2$ have
been implemented. Keeping the terms up to those linear in $l_2$, we get
\begin{eqnarray}
J_3&\approx&\frac{i}{4\pi^2}\int_0^1 du\int_0^{1-u}dv(1-u-v)\frac{2u(1-u)p_\lambda p_\beta
-(1-2u)vl_{2\lambda}p_\beta+(1-2u)vp_\lambda l_{2\beta}+g_{\lambda\beta}(1-2u)vp\cdot l_2}
{[u(1-u)s+v(1-v)l_2^2-2uvp\cdot l_2]^2}\nonumber\\
&\approx&\frac{i}{4\pi^2}\left[\frac{2p_\lambda p_\beta}{s^2}\ln\frac{s}{l_2^2}
+(g_{\lambda\beta}p\cdot l_2-l_{2\lambda}p_\beta+p_\lambda l_{2\beta})\frac{1}{s l_2^2}
\right].
\end{eqnarray}
To reach the last expression, we have focused on the leading region with small $u$ and $v$,
i.e., on the terms linear in $u$ and $v$.

We then work out the integral over $l_2$ in the small $l_2$ region,
\begin{eqnarray}
J_2&=&\frac{i}{4\pi^2}\int \frac{d^4l_2}{(2\pi)^4}
\frac{{\rm tr}[ (\not p-\not l_2)\gamma^\lambda(1-\gamma_5)
\not l_2\gamma^\sigma(1-\gamma_5)]}{(p-l_2)^2l_2^2}
\left[\frac{2p_\lambda p_\beta}{s^2}\ln\frac{s}{l_2^2}
+(g_{\lambda\beta}p\cdot l_2-l_{2\lambda}p_\beta+p_\lambda l_{2\beta})\frac{1}{s l_2^2}
\right]\nonumber\\
&\approx&\frac{2i}{\pi^2}\int \frac{d^4l_2}{(2\pi)^4}\frac{p^\lambda l_2^\sigma
+l_2^\lambda p^\sigma-g^{\lambda\sigma}l_2\cdot p}
{s l_2^2}\left[\frac{2p\cdot l_2}{s}\frac{2p_\lambda p_\beta}{s^2}\ln\frac{s}{l_2^2}
+(g_{\lambda\beta}p\cdot l_2-l_{2\lambda}p_\beta+p_\lambda l_{2\beta})\frac{1}{s l_2^2}
\right],\label{i3}
\end{eqnarray}
where the expansion of the denominator $(p-l_2)^2+i\epsilon\approx p^2-2p\cdot l_2+i\epsilon$
in powers of $l_2$ has been made. The above integral reduces to 
\begin{eqnarray}
J_2&\approx&\frac{2i}{\pi^2}\int \frac{d^4l_2}{(2\pi)^4}
\frac{1}{s^2}\left[\frac{4l_2^\sigma p\cdot l_2 p_\beta}
{s l_2^2}\ln\frac{s}{l_2^2}
+\frac{s l_2^\sigma l_{2\beta}-l_2^2p^\sigma p_\beta+p\cdot l_2 p^\sigma l_{2\beta}
+p\cdot l_2 l_2^\sigma p_\beta-g^\sigma_\beta (p\cdot l_2)^2}{(l_2^2)^2}\right]\nonumber\\
&\approx &\frac{3}{64\pi^4}\frac{p^\sigma p_\beta}{s}.
\end{eqnarray}
Note that the Wick rotation of the zeroth component of $l_2$ introduces $-i$,
instead of $+i$, since the pole is located in the first quadrant as indicated by the
denominator $p^2-2p\cdot l_2+i\epsilon$ in Eq.~(\ref{i3}). We have restricted the range of
the integration variable to $0<l_2^2<s/2$ after the Wick rotation in accordance with 
the soft approximation.

Finally, we group all the ingredients together, remove the lepton currents, and contract 
the remaining part by $g^{\alpha\beta}/4$, deriving 
\begin{eqnarray}
{\cal M}\approx\frac{2i}{8^4}\frac{m_{\mathcal{L}}^4 m_\ell^4}{v^8}
\frac{g^{\alpha\beta}}{4}\frac{i}{\pi^2}\frac{p_\alpha p_\sigma}{s^2} 
\frac{3}{64\pi^4}\frac{p^\sigma p_\beta}{s}
=-\frac{3}{128}\left(\frac{1}{16\pi^2}\right)^3
\frac{m_{\mathcal{L}}^4 m_\ell^4}{v^8}\frac{1}{s}.
\end{eqnarray}
The combination with the contribution from the diagram with horizontal scalar lines
then produces Eq.~(\ref{m22}).



\begin{thebibliography}{99}




\bibitem{Li:2023ncg}
H.~n.~Li,
Dispersive determination of neutrino mass ordering,
arXiv:2306.03463 [hep-ph].


\bibitem{Li:2023dqi}
H.~n.~Li,
Dispersive constraints on fermion masses,
Phys. Rev. D 107 (2023) 094007.


\bibitem{Li:2023yay}
H.~n.~Li,
Dispersive determination of electroweak-scale masses,
Phys. Rev. D 108 (2023) 054020.

\bibitem{Chien:2018ohd}
Y.~T.~Chien and H.~n.~Li,
Factorization of standard model cross sections at ultrahigh energy,
Phys. Rev. D 97 (2018) 053006.

\bibitem{Huang:2020iya}
L.~Huang, S.~D.~Lane, I.~M.~Lewis and Z.~Liu,
Electroweak Restoration at the LHC and Beyond: The $Vh$ Channel,
Phys. Rev. D 103 (2021) 053007.

\bibitem{Kaplan:1983fs}
D.~B.~Kaplan and H.~Georgi,
$SU(2)\times U(1)$ Breaking by Vacuum Misalignment,
Phys. Lett. B 136 (1984) 183.

\bibitem{PDG}
R.L. Workman et al. (Particle Data Group), Prog. Theor. Exp. Phys. (2022) 083C01.


\bibitem{Harrison:2002et}
P.~F.~Harrison and W.~G.~Scott,
$mu$-$tau$ reflection symmetry in lepton mixing and neutrino oscillations,
Phys. Lett. B 547 (2002) 219-228.

\bibitem{Alvarado:2020lcz}
J.~S.~Alvarado and R.~Martinez,
PMNS matrix in a non-universal $U(1)_{X}$ extension to the MSSM with one massless neutrino,
arXiv:2007.14519 [hep-ph].

\bibitem{Xu:2023kfi}
G.~Xu and Y.~Zhang,
CKM and PMNS mixing matrixes from $SO(2)$ flavor symmetry,
EPL 143 (2023) 44001.

\bibitem{Patel:2023qtw}
A.~A.~Patel and T.~P.~Singh,
CKM Matrix Parameters from the Exceptional Jordan Algebra,
Universe 9 (2023) 440.

\bibitem{Bora:2023teg}
H.~Bora, N.~K.~Francis, A.~Barman and B.~Thapa,
Neutrino mass model in the context of $\ensuremath{\Delta}(54)\ensuremath{\otimes}
Z_2\ensuremath{\otimes} Z_3\ensuremath{\otimes} Z_4$ flavor 
symmetries with Inverse Seesaw mechanism,
Phys. Lett. B 848 (2024) 138329.

\bibitem{Thapa:2023fxu}
B.~Thapa, S.~Barman, S.~Bora and N.~K.~Francis,
A minimal inverse seesaw model with S$_{4}$ flavour symmetry,
JHEP 11 (2023) 154.


\bibitem{Chung:2023rie}
Y.~Chung,
Dynamical origin of Type-I Seesaw with large mixing,
arXiv:2311.17183 [hep-ph].


\bibitem{Supanyo:2023jkh}
S.~Supanyo, C.~Hasuwannakit, S.~Yoo-Kong, L.~Tannukij and M.~Tanasittikosol,
The natural smallness of Dirac neutrino mass from the multiplicative Lagrangian,
arXiv:2312.16587 [hep-ph].

\bibitem{Lampe:2024gqo}
B.~Lampe,
Determination of Quark and Lepton Masses and Mixings in the Microscopic Model,
PoS EPS-HEP2023 (2024) 373.


\bibitem{Shaikh:2024ufv}
A.~R.~Shaikh and R.~Adhikari,
Explaining Fermions Mass and Mixing Hierarchies through $U(1)_X$ and $Z_2$ Symmetries,
arXiv:2404.11570 [hep-ph].


\bibitem{Bilenky:2002aw}
S.~M.~Bilenky, C.~Giunti, J.~A.~Grifols and E.~Masso,
Absolute values of neutrino masses: Status and prospects,
Phys. Rept. 379 (2003) 69.

\bibitem{see}
P.~Minkowski,
$\mu \to e\gamma$ at a Rate of One Out of $10^{9}$ Muon Decays?,
Phys. Lett. B 67 (1977) 421-428;
M.~Gell-Mann, P.~Ramond and R.~Slansky,
Complex Spinors and Unified Theories,
Conf. Proc. C 790927 (1979) 315-321;
T.~Yanagida,
Horizontal gauge symmetry and masses of neutrinos,
Conf. Proc. C 7902131 (1979) 95-99;
R.~N.~Mohapatra and G.~Senjanovic,
Neutrino Mass and Spontaneous Parity Nonconservation,
Phys. Rev. Lett. 44 (1980) 912.


\bibitem{Nature}
The KATRIN Collaboration, 
Direct neutrino-mass measurement with sub-electronvolt sensitivity,
Nat. Phys. 18 (2022) 160. 



\bibitem{Mohapatra:2006gs}
R.~N.~Mohapatra and A.~Y.~Smirnov,
Neutrino Mass and New Physics,
Ann. Rev. Nucl. Part. Sci. 56 (2006) 569;
R.~N.~Mohapatra,
Neutrino mass as a signal of TeV scale physics,
Nucl. Phys. B 908 (2016) 423.

\bibitem{Li:2022jxc}
H.~n.~Li,
Dispersive analysis of neutral meson mixing,
Phys. Rev. D 107 (2023) 054023.

\bibitem{Li:2020xrz} 
H.~n.~Li, H.~Umeeda, F.~Xu and F.~S.~Yu,
$D$ meson mixing as an inverse problem,
Phys. Lett. B 810 (2020) 135802.

\bibitem{Cheng} 
H.~Y.~Cheng,
CP Violating Effects in Heavy Meson Systems,
Phys. Rev. D 26 (1982) 143.

\bibitem{BSS}
A.~J.~Buras, W.~Slominski and H.~Steger,
$B^0$ anti-$B^0$ Mixing, CP Violation and the $B$ Meson Decay,
Nucl. Phys. B 245 (1984) 369-398.


\bibitem{deSalas:2017kay}
P.~F.~de Salas, D.~V.~Forero, C.~A.~Ternes, M.~Tortola and J.~W.~F.~Valle,
Status of neutrino oscillations 2018: 3$\sigma$ hint for normal mass ordering and improved 
CP sensitivity,
Phys. Lett. B 782 (2018) 633.

\bibitem{Planck:2018vyg}
N.~Aghanim \textit{et al.} [Planck],
Planck 2018 results. VI. Cosmological parameters,
Astron. Astrophys. 641 (2020) A6
[erratum: Astron. Astrophys. 652 (2021) C4].


\bibitem{Alvey:2021sji}
J.~Alvey, M.~Escudero and N.~Sabti,
What can CMB observations tell us about the neutrino distribution function?,
JCAP 02 (2022) 037.


\bibitem{Long:2022dil}
H.~Long, C.~Morales-Guti\'errez, P.~Montero-Camacho and C.~M.~Hirata,
Impact of inhomogeneous reionization on post-reionization 21-cm intensity mapping measurement 
of cosmological parameters,
Mon. Not. Roy. Astron. Soc. 525 (2023) 6036-6049.


\bibitem{DESI:2024mwx}
A.~G.~Adame \textit{et al.} [DESI],
DESI 2024 VI: Cosmological Constraints from the Measurements of Baryon Acoustic Oscillations,
arXiv:2404.03002 [astro-ph.CO].

\bibitem{CK84}
L.~L.~Chau and W.~Y.~Keung,
Comments on the Parametrization of the Kobayashi-Maskawa Matrix,
Phys. Rev. Lett. 53 (1984) 1802;
L.~Maiani,
New Currents,
in Proceedings of the 8th International Symposium on Lepton and 
Photon Interactions at High Energies, DESY, Hamburgh, 1977, pp. 867-894.






\bibitem{Buttazzo:2013uya}
D.~Buttazzo, G.~Degrassi, P.~P.~Giardino, G.~F.~Giudice, F.~Sala, A.~Salvio and A.~Strumia,
Investigating the near-criticality of the Higgs boson,
JHEP 12 (2013) 089.



\end{thebibliography}
\end{document}